\documentclass[11pt]{article}
\usepackage{amsbsy}
\usepackage{amssymb}
\usepackage{graphicx}
\usepackage{url}
\usepackage{hyperref}
\usepackage[utf8]{inputenc}
\usepackage{color}
\usepackage{amssymb}\usepackage{epsfig}
\let\oldbibliography\thebibliography
\renewcommand{\thebibliography}[1]{%
  \oldbibliography{#1}%
  \setlength{\itemsep}{2pt}%
}
\newcommand{\BE}{\begin{equation}}
\newcommand{\EE}{\end{equation}}
\newcommand{\half}{\frac{1}{2}}
\usepackage[left=2.5cm,right=2.5cm,top=2.7cm,bottom=2.7cm]{geometry}
\begin{document}
\begin{titlepage}
\begin{center}

\vskip 1 pt

{\LARGE\bf The 690 GeV scalar resonance}

\end{center}

\begin{center}

\vspace*{14mm}{\Large M.~Consoli$^{(a)}$, L.~Cosmai$^{(b)}$, F.~Fabbri$^{(c)}$,
and G.~Rupp$^{(d)}$}
\vspace*{4mm}\\
{a) Istituto Nazionale di Fisica Nucleare, Sezione di Catania, Italy ~~~~~\\
 b) Istituto Nazionale di Fisica Nucleare, Sezione di Bari, Italy ~~~~~~~~~\\
 c) Istituto Nazionale di Fisica Nucleare, Sezione di Bologna, Italy ~~~\\
 d) CFTP, Instituto Superior T\'{e}cnico, Universidade de Lisboa, Lisboa,
Portugal}
\end{center}

\begin{center}
{\bf Abstract}
\end{center}
\par \noindent
Spontaneous symmetry breaking through the Higgs field has been experimentally
confirmed as a basic ingredient of the Standard Model. However, the origin of
the phenomenon may not be entirely clear, because, in perturbation theory,
the vacuum turns out to be a metastable state. An alternative scenario was
proposed that implies a second resonance of the Higgs field  ${\cal H}$ with a
well delimited mass $(M_H)^{\rm Theor} = 690\,(30)$~GeV. This stabilises
the potential, but, owing to an $H$ coupling to longitudinal $W$s with the
same typical strength as that of the low-mass state with $m_h= 125$~GeV, it
would still remain a relatively narrow resonance. Our scope here is twofold.
First, leaving out many details, we outline a simple logical path where the,
apparently surprising, idea of such a second resonance follows from
basic properties of $\Phi^4$ theories. Secondly, we spell out a definite
experimental signature of this resonance that is clearly
visible in various LHC data. As a by-product, the ${\cal H} ^3$ term
gives $\kappa_\lambda = (M_H/m_h) \sim $ 5.5 consistently with the ATLAS
and CMS data.

%}

%\end{center}

%\vskip 10 pt \par\noindent PACS: 03.30.+p; 01.55.+b; 11.30.Cp

%\end{center}

%\vskip 10 pt \par\noindent PACS: 03.30.+p; 01.55.+b; 11.30.Cp
\end{titlepage}

\section{Introduction}

Extensive LHC data have shown that the scalar resonance with
mass $m_h=125$ GeV  \cite{discovery1,discovery2}  couples to the other particles proportionally to their
respective masses. Spontaneous symmetry breaking (SSB), through the Higgs
field, has thus been experimentally confirmed as a basic ingredient of the
Standard Model (SM). However, the physical origin of the phenomenon may not
be entirely clear. Indeed, within perturbation theory, the scalar self-coupling
$\lambda^{\rm (p)}(\phi)$ (p=perturbative) starts to decrease from its value
$\lambda^{\rm (p)}(v)=3m^2_h/v^2$ at the Fermi scale $v\sim 246$~GeV, becoming
negative beyond an instability scale $\phi_{\rm inst} \sim 10^{10}$~GeV.
The true minimum of the perturbative large-$\phi$ potential
$V^{(p)}(\phi) \sim \lambda^{(p)}(\phi)\phi^4$ would then lie beyond the
Planck scale and be much deeper than the electroweak minimum, implying our
vacuum to be a metastable state.\footnote{While new physics, near or below the
Planck scale \cite{branchina,gabrielli}, may substantially modify the whole
perspective, the issue of metastability within the SM, with the increasing
precision of measurements, has been more and more refined theoretically.
To that end, we refer to \cite{DESY}, which presents the state of the art
for this problem. In particular, the critical top-quark mass for stability
up to the Planck mass was found to be consistent, within 1.3$\sigma$, with the
corresponding value of the pole mass from Monte Carlo simulations. This
suggests that the familiar idea of metastability may not yet be definitely
settled.}

In this perspective, it was proposed \cite{Cosmai2020}$-$\cite{EPJC} to first
restrict to the pure scalar sector and describe SSB as in those studies in
which the mass $m_h$ defining the quadratic shape of the potential at its
minimum is much smaller than the mass $M_H$ associated with the zero-point
energy (ZPE) determining the potential depth. Therefore, differently from
perturbation theory, where these two mass scales coincide, the Higgs field
could exhibit a second resonance. In the present paper, our scope is
twofold. In the first place, important yet technical details have probably
obscured the physical motivation for a second resonance, which, as we will
show, directly follows from basic properties of $\Phi^4$ theories. Secondly, we
point out a characteristic, experimental signature of this resonance that is
clearly visible in several LHC data.

\section{Second resonance: why 690 GeV}

A non-perturbative description of SSB can hardly be obtained within
the full gauge and fermion structure of the theory. Thus, one could first
restrict to the pure scalar sector and check, a posteriori, 
that the other couplings do not play a significant role. Concerning pure
$\Phi^4$ theories, analytical and numerical studies require:
i) a continuum limit with a Gaussian structure of Green's functions
(``triviality'') and ii) a picture of SSB as a weak first-order
(or quasi-first-order) phase
transition.\footnote{As for lattice simulations, one can just look at
fig.~7 in \cite{akiyama2019phase}, where the data for the average field
near the critical temperature show the characteristic first-order jump and not
a smooth second-order trend.}.
We are thus lead to consider Gaussian-type forms of the effective potential,   
encompassing some classical background plus the ZPE of free-field-like
fluctuations with a $\phi$-dependent mass. SSB would then originate from the
ZPE in the classically scale-invariant limit $ V''_{\rm eff}(\phi=0)\to 0^+$,
as in the original Coleman-Weinberg paper \cite{Coleman:1973jx}, which is the
prototype of such calculations. 

\begin{figure}[ht]
\centering
\includegraphics[width=0.45\textwidth,clip]{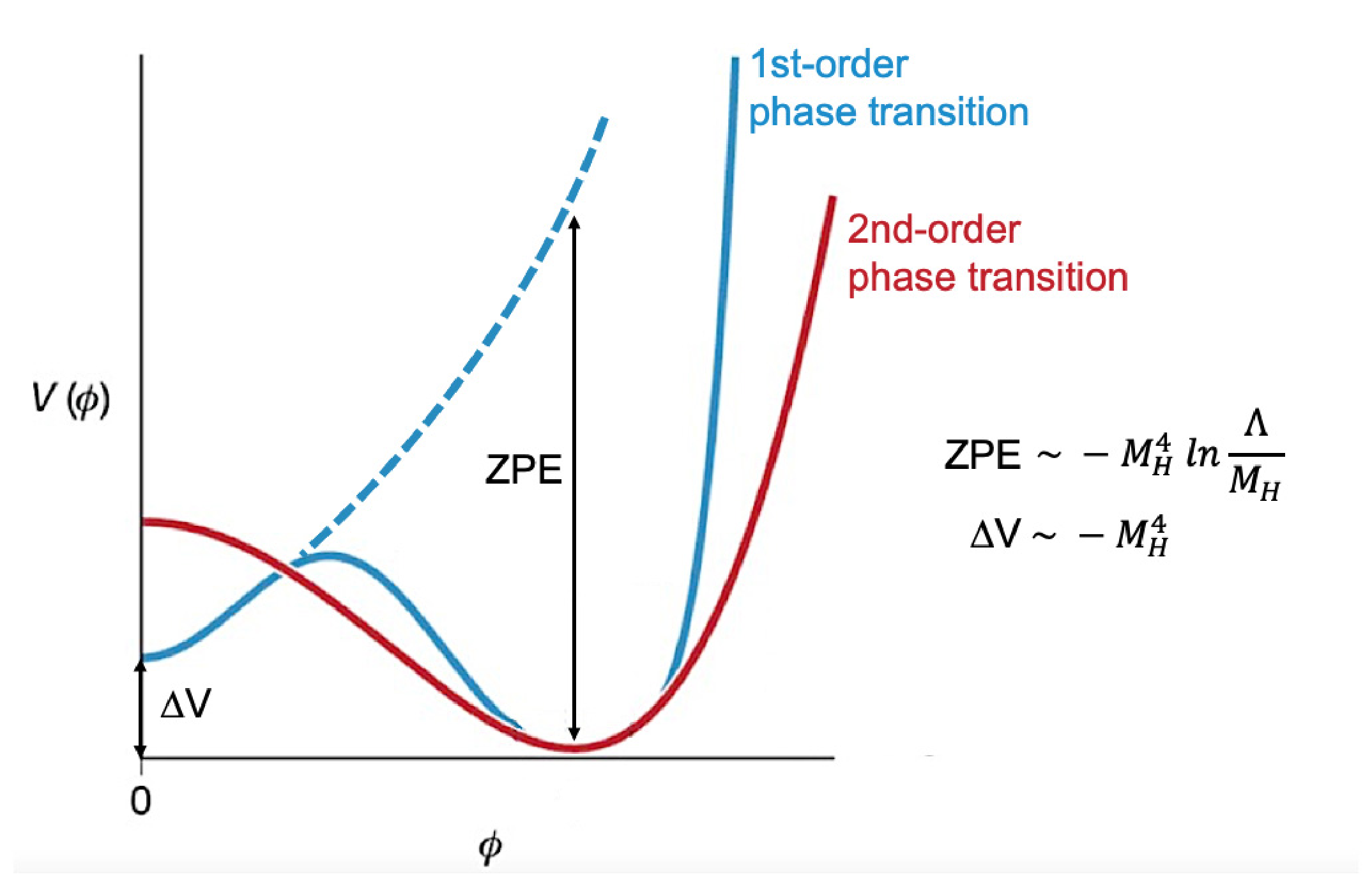}
\caption{A picture illustrating the role of the ZPE in a 
first-order scenario of SSB. Differently from the standard second-order
picture, this has to compensate for a tree-level potential with no
non-trivial minimum.}
\label{zpe}
\end{figure}

This first-order scenario could now explain the existence of two mass scales;
see fig.~\ref{zpe}.
The red and blue curves have the same quadratic shape at the minimum,
say $m^2_h$.  But, differently from a second-order picture, where SSB
originates in a negative quadratic curvature at $\phi=0$, the ZPE is now
overwhelming a tree-level potential with no non-trivial minimum. Thus,
the ZPE mass scale $M_H$ and the mass scale $m_h$ play different roles.
This difference finds its quantitative description in a Renormalisation
Group (RG) analysis of the effective potential, where the two masses scale
differently with the ultraviolet cutoff $\Lambda$ placed by the Landau pole
of $\Phi^4$. There are then two $\Lambda$-independent quantities, namely $M_H$
itself, entering the minimum of the effective potential $\Delta V \sim -M^4_H$,
and a particular definition of the vacuum field to be used for the Fermi scale
$v\sim 246$~GeV, which is the relevant one for the gauge and fermion fields
and is always assumed to be cutoff independent. As such, $M_H$ and $v$ can be
related through a simple proportionality constant, say $M_H=Kv$.
Instead, for the smaller mass $m_h$, defining the quadratic shape of the
potential, i.e., the inverse zero-momentum propagator $G^{-1}(p=0)= m^2_h$,
one finds $m^2_h \sim  M^2_H  L^{-1}\sim v^2 L^{-1} $ in terms of
$L=\ln (\Lambda/M_H)\sim \ln (\Lambda/v)$, thus implying the traditional
$\Phi^4$ relation $\lambda(v)=3m^2_h/v^2\sim L^{-1}$. 

The same two-mass structure, is obtained from the Gaussian Effective Action
(GEA), both for the one-component and $O(N)$-symmetric theory \cite{okopinska},
which gives
\BE
G^{-1}(p) \; = \; p^2 +  M^2_H A(p) \; .
\label{GEA_general}
\EE
In fact, upon minimisation of the Gaussian effective potential, one obtains
\cite{EPJC}
$G^{-1}_h(p)\sim  p^2 + m^2_h$ for $p \to 0$, where $A(p)\sim  L^{-1}$, and
$G^{-1}_H(p)\sim p^2 + M^2_H$ at larger $p^2$, where $A(p) \sim 1$. 
Equation~(\ref{GEA_general}) has also been checked with lattice
simulations \cite{Cosmai2020} of the scalar propagator. These
simulations were also needed because those Gaussian-like approximations to
the effective potential predict the same qualitative scaling pattern, but,
resumming to all orders different classes of diagrams,
yield different values of the numerical coefficient $c_2$ controlling the
logarithmic slope, say $M_H = m_h (L/c_2)^{1/2}$. Therefore, fitting the
lattice propagator to a free-field form, one could compute $m_h$ from the
$p\to 0$ limit of $G(p)$ and also  $M_H$ from its behaviour at higher
$p^2$. Referring to \cite{Cosmai2020,EPJC} for details, the expected
logarithmic trend was checked with the result $(c_2)^{-1/2} = 0.67\,(3)$. 
Therefore, by combining with $m^2_h = \lambda (v) v^2/3$ and the leading-order
trend $\lambda(v)\sim (16 \pi^2/3)  L^{-1}$ of $\Phi^4$, giving
$m_h \sim (4\pi v/3) \cdot L^{-1/2} $, one could find the finite
proportionality constant $K= (4\pi/3) \cdot(c_2)^{-1/2} $, or
\BE
(M_H)^{\rm Theor} \; = \; Kv= 690\,(30) \; {\rm GeV} \; .
\label{MHTHEOR}
\EE
Note that here one finds $m_h \ll M_H$ for very large
$\Lambda$. But $M_H$ is independent of $\Lambda$, so that by decreasing
$\Lambda$ the lower mass will increase and
approach its maximum value $(m_h)^{\rm max}\sim M_H \sim 690\,(30)$~GeV when 
$\Lambda$ becomes as small as possible, say a few times $M_H$. Then, by
comparing this prediction with the known upper bounds from lattice simulations,
we find good consistency with Lang's \cite{lang} and Heller's \cite{heller}
values, viz.\ $(m_h)^{\rm max}=670\,(80)$~GeV and
$(m_h)^{\rm max}=710\,(60)$~GeV, respectively. Actually, the combination of
these two estimates, say $(m_h)^{\rm max} \sim 690\,(50)$~GeV, would
practically coincide with our expectation. This remarkable agreement confirms
that, if a second resonance exists, its mass can only be in a narrow window
around 690 GeV.  At the same time, in the real world $m_h=125$~GeV.
Therefore, if a second resonance with $M_H\sim 690$~GeV exists,
the ultraviolet cutoff $\Lambda$ must be extremely large. 

\section{Second resonance: basic phenomenology}

The preceding discussion concerned the description of SSB in the pure scalar
sector. We will now consider the other parts of the theory, by restricting
ourselves to the observable implications at the Fermi scale.  As a first
remark, we observe that, with
such a large mass $M_H\sim 690$~GeV in the scalar ZPE, the known gauge
and fermion fields would play a minor role. In fact, the logarithmically
divergent terms in the ZPE are proportional to the fourth power of the mass,
so that in units of the pure scalar term one finds
$(6 M^4_w + 3 M^4_Z)/M^4_H \lesssim 0.002$ and $12 m^4_t/M^4_H\lesssim 0.05$.
Moreover, the two couplings $\lambda^{\rm (p)}(\mu)$ and  $\lambda(\mu)$
coincide for $\mu=v$, which implies that their large-$\mu$ difference
remains unobservable. Confirming this alternative mechanism of SSB would then
require observing the second resonance and checking its phenomenology.

Here, the hypothetical $H$ differs from a conventional Higgs boson
of 700~GeV, because its large conventional width to longitudinal $W$s is
suppressed by the ratio $(m_h/M_H)^2 \sim 0.032$ \cite{memorial,EPJC}. 
This suppression effect can be deduced on the basis of two
different arguments. First, one should critically consider the tree-level
calculations in the unitary gauge. These show that, for a large $M_H$,
longitudinal $W$s effectively interact at high energy as in a $\Phi^4$ theory
with contact coupling $\lambda_0 \sim 3(M^2_H/v^2)\gg 1$. The appearance of
this coupling, which is not present in the vector-boson Lagrangian, indicates
that higher-order $W_L W_L$ rescattering is far from negligible and should
be included, for instance, to obtain a reliable estimate of the $H\to WW$
decay width. In the model of an effective $\Phi^4$ theory with cutoff
$\Lambda$, these higher-order effects would then produce the replacement
$\lambda_0 \to \lambda(E)  \sim  [\ln(\Lambda/E)]^{-1}$. Therefore, to
logarithmic accuracy, such that $\ln (\Lambda/E) \sim \ln (\Lambda/M_H)$,
one may envisage the replacement $M^2_H/v^2 \to m^2_h/v^2$. A second, and
more precise derivation is instead obtained by considering the class of
renormalisable $R_\xi$ gauges with total Lagrangian 
\BE
\label{langrangian}
L \; = \; L_{\rm inv}-U_{\rm scalar}+L_{\rm GF}+L_{\rm FP}+L_{\rm fermion} \;. 
\EE
The invariant Lagrangian, given for simplicity for pure $SU(2)$ symmetry, with
gauge coupling $g_{\rm gauge}=g$ and $\sin \theta_w=0$, reads
\begin{eqnarray}
L_{\rm inv}  = -\frac{1}{4} G^a_{\mu\nu} G^a_{\mu\nu} +
{\half}{\cal H}G^{-1}{\cal H} - {\half} M^2_w W^2(1 + {\cal H}/v)^2 
-{\half} \partial_\mu \chi^a D_\mu \chi^a  \nonumber\\
+{\half} gW^a_\mu({\cal H}\partial_\mu \chi^a -\chi^a \partial_\mu {\cal H}) 
-M_w \chi^a \partial_\mu W^a_\mu -\frac{1}{8}g^2 W^2 \chi^a\chi^a
\label{lagrangian}
\end{eqnarray}
where the term $ {\half}{\cal H}G^{-1}{\cal H}$ replaces the one-pole structure ${\half}{\cal H}(\partial^2 - M^2_H){\cal H}$. Also, $L_{\rm GF}$ and
$L_{\rm FP}$ are the gauge-fixing and Faddeev-Popov terms, respectively,
while $L_{\rm fermion} $ has the usual form
$(1 + {\cal H}/v) \sum_f m_f \bar f f $, after symmetry breaking. 

Replacing the term $ {\half}{\cal H}G^{-1}{\cal H}$ corresponds to the resummation of all one-loop bubbles, in the scalar sector, which is operated by the GEA 
 (see \cite{EPJC} for a diagrammatic representation). Instead, by the same resummations, at the minimum of the effective potential, the $\chi^a$ fields remain massless as in perturbation theory \cite{okopinska}. In this way, the scalar sector is treated non-perturbatively while the effects of the gauge couplings can be added perturbatively. This different treatment of the two sectors is in the spirit of an effective Lagrangian and is consistent with the content of the Equivalence Theorem in renormalizable $R_\xi$ gauges that holds to lowest order in $g$ and to all orders in the scalar self interactions \cite{bagger}. Therefore, at high energy, up to $O(M_w/E)$ terms, the results for the pure scalar sector are only modified by $O(g^2)$ terms and one can write down a relation between two different $S$-matrix elements in terms of some constant $C=1+ O(g^2)$
\BE
\label{redux} 
S[W_L's ,{\rm physical}] = C^n \tilde S[\chi's,{\rm physical}] \; .
\EE 
The left-hand side indicates the matrix element, with $n$ longitudinal $W$s
plus an arbitrary number of other physical states, in the full theory
where $g\neq 0$. The right-hand side refers instead to the corresponding
matrix element in the theory at $g=0$, where the Goldstone bosons are now
physical states. In particular, for the decay $H\to WW$, the S-matrix element $S[H, W_L,W_L]$ can be expressed in terms of the corresponding matrix element $\tilde S[H,\chi,\chi]$ at $g=0$ up to small corrections. To fully appreciate the implications, let us consider the structure
of the Higgs potential in terms of two parameters $\epsilon_1$ and
$\epsilon_2$ \cite{memorial}, i.e.,
\BE
\label{scalar}
U_{\rm scalar} \; = \; \epsilon_1 \frac{M^2_H}{2v } {\cal H} ( {\cal H}^2 +
\chi^a\chi^a) + \epsilon_2 \frac{M^2_H}{8v^2} ({\cal H}^2+\chi^a\chi^a)^2 \; .
\EE
By setting $\epsilon_1=1$, the resulting width
$\Gamma(H\to WW) \sim (1 + O(g^2)) \epsilon^2_1 (M^3_H/v^2)$ gives the
conventional result $\Gamma^{\rm conv}(H\to WW) \sim G_F M^3_H$. However, if
the vacuum field $v\sim 246$~GeV scales uniformly with $M_H$, both $M^2_H/v$
and $M^2_H/v^2$ would be $\Lambda$-independent. As such, by ``triviality'',
they cannot represent the interaction terms far below the ultraviolet cutoff.
This is why, by minimizing the effective potential in those Gaussian-like approximations, one finds
$\epsilon_1=(m_h/M_H) \sim L^{-1/2}$ and $\epsilon_2=\epsilon^2_1 \sim L^{-1}$.
The resulting four-linear coupling $\epsilon_2 (M^2_H/8v^2)= (m^2_h/8v^2)$ is
then the same as in perturbation theory, whereas, from the cubic term, we get
the anticipated suppression of the decay width
$\Gamma(H\to WW+ ZZ) \sim 0.032 \times \Gamma^{\rm conv}(H\to WW+ ZZ)$.

Note that the ${\cal H}^3$ term $\epsilon_1(M^2_H)/(2v)=(m_hM_H)/(2v)$ is smaller than $M^2_H/(2v)$, but also larger than its perturbative value $m^2_h/(2v)$. This difference can thus be reabsorbed into a rescaling
$\kappa_\lambda$ as
\BE 
\label{kappalambda}
\kappa_\lambda \; = \; \frac{M_H}{m_h }=(690 \pm 30)/125=5.52 \pm 0.24 \;. 
\EE
In spite of this difference, eq.~(\ref{kappalambda}) is consistent with the
recent ATLAS and CMS determinations (see Fig.~3c of \cite{atlascouplings} and Figs.~5 and 7 of \cite{CMScouplings}) and might actually explain the rather asymmetric
shapes of the observed probability contours with best-fit values
$\langle\kappa_\lambda\rangle^{\rm exp}=4.3$ and 4.7, respectively. 

In conclusion, by setting $ G^{-1}=(\partial^2 -M^2_H)$ and $\epsilon_1=\epsilon_2=1$, we would get the usual theory with a single mass $M_H$. Here, instead, we are faced with a two-mass structure and the following numerical estimates
$\Gamma(H \to ZZ)\sim \frac { M_H} { 700~ {\rm GeV}}\,\times\,$(1.6~GeV),
$\Gamma(H\to WW)\sim \frac { M_H} { 700~ {\rm GeV}}\,\times\,$(3.3 GeV),
$\Gamma(H \to hh)\sim \frac { M_H} { 700~ {\rm GeV} }\,\times\,$(1.5 GeV).
As such, the heavy $H$ should be a relatively narrow resonance of total width $\Gamma(H\to all) =25\div35$~GeV, decaying predominantly to $t \bar t$ quarks, with a branching ratio of about 75$\div$80 $\%$.  Note the very
close branching ratios $B(H \to hh) \sim 0.94~ B(H \to ZZ)$. Finally, due to its small coupling to longitudinal $W$s, $H$ production through vector-boson fusion (VBF) would be negligible as compared to gluon-gluon fusion
(ggF), which has a typical cross section
$\sigma^{\rm ggF} (pp\to H) = 1000\,(200)$~fb ,
for $M_H$ in the range $690\,(30)$~GeV.

\section{Second resonance: a characteristic experimental signature}
\begin{figure}[ht]
\centering
\includegraphics[width=0.45\textwidth,clip]{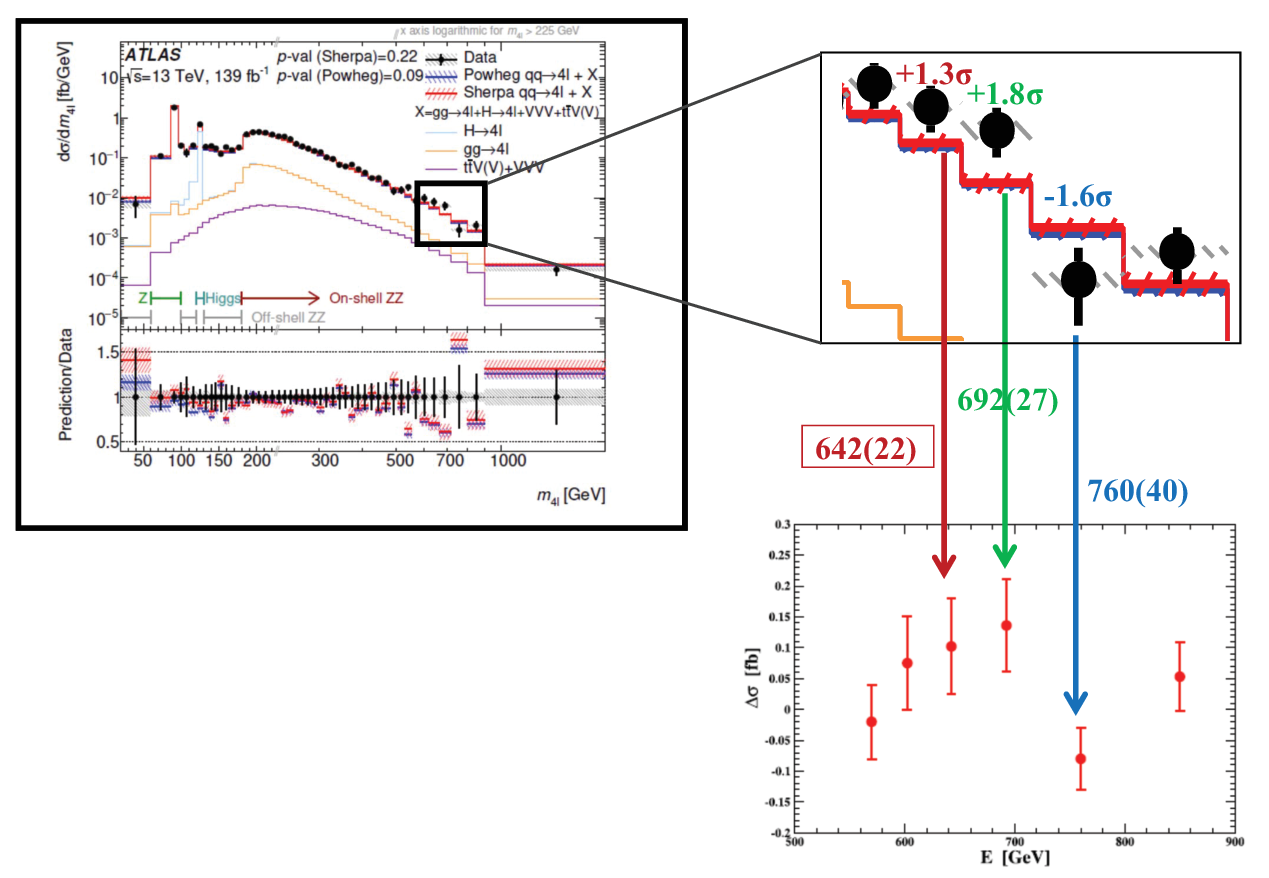}
\caption{The average differential cross section measured by ATLAS
\cite{atlasnew}, in bins of increasing size, and the difference
of measured and background cross sections as reported in
table~\ref{leptonxsection}. }
\label{ATLAS_cross_section}
\end{figure}

In a first contact with experiment, given the expected large branching ratio
$B(H\to t\bar t)=(75\div 80)\%$, the most natural place to look for 
the new resonance is in the $t\bar t$ channel. However, in
the relevant region of invariant mass $m(t\bar t)=620\div820$~GeV,
CMS measurements \cite{CMS_top} give a background cross section
$\sigma(pp\to t\bar t)=107\pm7.6$~pb that is about 100 times
larger than the expected signal $\sigma(pp\to H \to t \bar t) \lesssim 1$~pb.
For this reason, it is remarkable that a slight excess has been observed
\cite{ATLAS_conf} in our region of invariant mass, viz.\ around 675 GeV,
especially for those events where the tracks of the two final leptons are at
large angles; see fig.~12 of \cite{ATLAS_conf}. The excess is minuscule, just
a $1\%$ effect, but {\it this is precisely the expected value} for a 1~pb
signal on a 100~pb background. Therefore, it represents an interesting
indication, even though we are speaking of a $1\sigma$ effect.
\small
\begin{table*}
\caption{The observed ATLAS \cite{atlasnew} 4-lepton cross section and
the estimated background in the range of invariant mass $m(4l)\equiv E$ from
555 to 900~GeV. These values have been obtained by multiplying the bin size
with the average differential cross sections $\langle (d \sigma/d E)\rangle$,
reported for each bin in the companion HEPData file.}.
\begin{center}
\begin{tabular}{cccc}
Bin [GeV]& ${\sigma}_{\rm EXP} $~[fb]  & $\rm{ \sigma}_{\rm B} $~[fb] &
$(\sigma_{\rm EXP} - \rm {\sigma}_{\rm B} $)~[fb]  \\
\hline \hline
555--585 & 0.252 $\pm 0.056$ & $0.272 \pm 0.023$ & $-0.020 \pm 0.060$  \\
\hline
585--620 & $0.344\pm 0.070$ & $ 0.259 \pm 0.021 $ & $+0.085 \pm 0.075$ \\
\hline
620--665 & $0.356 \pm 0.075$ & $ 0.254 \pm 0.023$ & $+0.102\pm 0.078$  \\
\hline
665--720 & $0.350 \pm 0.073$ &$ 0.214 \pm 0.019$  &$+0.136 \pm 0.075$ \\
\hline
720--800 & $0.126 \pm 0.047$ &$ 0.206 \pm 0.018$ & $ -0.080 \pm 0.050$ \\
\hline
800--900 & $0.205\pm 0.052$ &$ 0.152 \pm 0.017$  & $+0.053 \pm 0.055$  \\
\hline
\end{tabular}
\end{center}
\label{leptonxsection}
\end{table*}
\normalsize
We will now consider the decay $H\to ZZ$ and postpone the $H\to hh$ at the end of this chapter. Here, for $B(H\to ZZ)=0.053\,(12)$, $\sigma^{\rm ggF} (pp\to H) = 1000\,(200)$~fb and
$4B^2(Z\to l^+l^-)\sim$ 0.0045, we expect a peak cross section 
$\sigma_{\rm peak}(pp\to H\to 4l)=0.25\,(10)$~fb. We can thus compare with the
ATLAS cross section in the charged 4-lepton channel; see
table~\ref{leptonxsection} and fig.~\ref{ATLAS_cross_section}. These ATLAS
determinations, in the relevant region $665\div720$~GeV, indicate an average background cross section $\langle \sigma_B\rangle=0.214$~fb, so of the same magnitude as the expected Breit-Wigner (BW) peak. 
For $\sigma_{\rm peak}(pp\to H\to 4l)=0.25\,(10)$~fb we then find a ratio 
\BE
\frac{\sigma_{\rm peak} } {\sigma_{\rm B} } = 0.7 \div 1.6  .
\EE
Of course this is just an order of magnitude. To compare with the data, we
should also consider background/resonance interference and the effect of data
binning. 
\begin{figure}[ht]
\centering
\includegraphics[width=0.30\textwidth,clip]{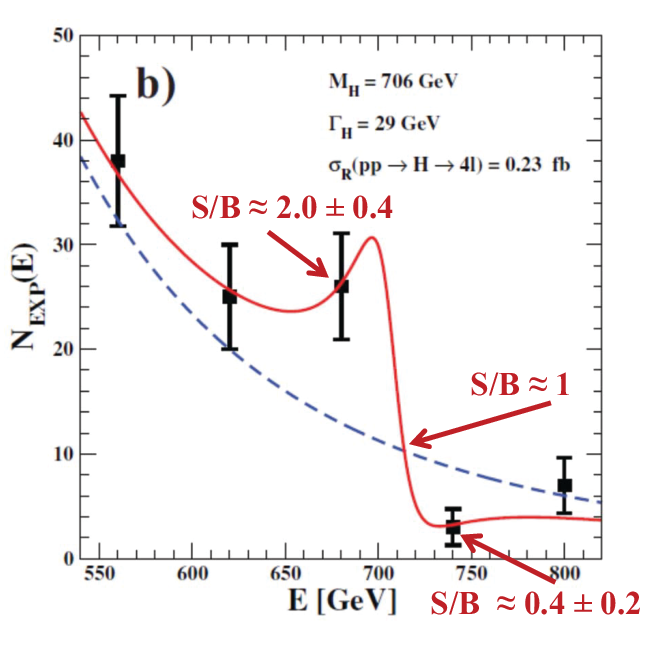}
\caption{The dominant ATLAS ggF-low 4-lepton events \cite{atlas4lHEPData}
in bins of 60~GeV from 530 to 830~GeV. The integrated luminosity is
139~fb$^{-1}$ and the average acceptance is 0.38. The red curve is the fit
with eq.~(\ref{sigmat}) and the dashed blue curve is the background.}
\label{ATLAS_ggF_low}
\end{figure}
 \small 
\begin{table*}
\caption{Values
of $S/B = \sigma_T/ \sigma_B$  eq.~(\ref{sigmat}) in bins of size $\Delta$, 
for $\Gamma/M=0.03$ and
$\langle\frac{\sigma_{\rm peak}}{\sigma_{\rm B} }\rangle=0.7$.}
\begin{center}
\begin{tabular}{crc}
$\Delta=\Gamma$      &   $\Delta=2\Gamma$  &  $\Delta=2\Gamma$   \\
\hline \hline
$[M\!-\!3/2\,\Gamma , M\!-\!1/2\,\Gamma]$: $S/B = 1.80$ &
$[M\!-\!3\,\Gamma , M\!-\!\Gamma]$: $S/B = 1.44$ &
$[M\!-\!5/2\,\Gamma , M\!-\!1/2\,\Gamma]$: $S/B = 1.60$ \\
\hline
$[M\!-\!1/2\,\Gamma , M\!+\!1/2\,\Gamma]$: $S/B = 1.54$ &
$[M\!-\!\Gamma, M\!+\!\Gamma]$: $S/B = 1.39$ &
$[M\!-\!1/2\,\Gamma, M\!+\!3/2\,\Gamma]$: $S/B = 1.03$ \\
\hline
$[M\!+\!1/2\,\Gamma , M\!+\!3/2\,\Gamma]$: $S/B = 0.47$ &
$[M\!+\!\Gamma , M\!+\!3\,\Gamma]$: $S/B = 0.61$ &
$[M\!+\!3/2\,\Gamma , M\!+\!7/2\,\Gamma]$: $S/B = 0.65$ \\
\hline
\end{tabular}
\end{center}
\label{simulation}
\end{table*}
\begin{table*}[ht]
\caption{The ATLAS \cite{atlas4lHEPData} ggF-like 4-lepton events for the
various categories (see text) with, in parentheses, the expected background.
The error bars in the $S/B$ ratio reflect the $\sqrt{N}$ statistical
uncertainty of the observed events.}
\begin{center}
\begin{tabular}{ccccccc}
$\rm E$ [GeV] & high-4$\mu$ & high-2e2$\mu$ & high-4e & low &Total & S/B  \\
\hline
560\,(30) & 4 (3.6) & 3 (6.2) & 5 (2.7)  & 38 (32.0) &50 (44.5)&$1.13\pm0.16$\\
\hline
620\,(30) & 3 (2.3) & 2 (3.9) & 4 (1.7) & 25 (20.0)  &34 (27.9)&$1.22\pm0.21$\\
\hline
680\,(30) & 1 (1.5) & 2 (2.5) & 3 (1.1) & 26 (13.0)  &32 (18.1)&$1.77\pm0.31$\\
\hline
740\,(30) & 0 (1.0) & 1 (1.6) & 0 (0.7) &  3 (8.7) & 4 (12.0) & $0.33\pm0.17$\\
\hline
800\,(30) & 1 (0.7) & 2 (1.1) & 0 (0.5) & 7 (6.0) & 10 (8.3) & $1.21\pm0.38$\\
\hline
\end{tabular}
\end{center}
\label{ATLAS-MVA}
\end{table*}
\normalsize
To that end, we adopt the basic model of a resonating amplitude $A_R$ 
interfering with a real amplitude $A_B$ producing a background cross section $\sigma_B$ . By defining $s=E^2$
and $x= (M^2- s)/(M\Gamma)$, we find a total cross section 
\BE
\sigma_T \; = \; \sigma_B+\frac{\sigma_{\rm peak}}{1+x^2} +
\frac{2x}{1+ x^2} \sqrt{\sigma_B \sigma_{\rm peak}} \;.  
\label{sigmat}
\EE
At this very early stage, when we do not even know for certain whether there
is a resonance or not, this is a convenient, zeroth-order scheme to address
the data in a model-independent way. More refined estimates, at the
quark-parton level, which distinguish among the possible production
mechanisms, will enter at a next level of
precision.\footnote{After identifying the resonance with a model-independent
analysis of the data, this refinement will be needed if one assumes a
definite production mechanism at the quark-parton level, e.g.\ through ggF.
Then, denoting by $\sigma^{\rm gg}_B$ the specific 4-lepton 
background from the ggF mechanism given in \cite{atlasnew},
the ``non-ggF'' background was preliminarily subtracted in
\cite{LHEP,EPJC} by defining a modified experimental cross section
$\hat{\sigma}_{\rm EXP}\;=\;\sigma_{\rm EXP}-(\sigma_B-\sigma^{\rm gg}_B)$
and then replacing everywhere $\sigma_B\to\sigma^{\rm gg}_B$ in the
theoretical eq.~(\ref{sigmat}).}  
With eq.~(\ref{sigmat}), we thus consider the size of the interference and the
effect of binning the data in regions of width $\Delta$. The results of this
simulation in table~\ref{simulation} indicate that if the BW peak is
comparable to the background, an important signature
is provided by excess/defect sequences. In particular, the numbers in the
third column indicate that we could see no enhancement in the bin that
includes the mass.
\begin{figure}[ht]
\centering
\includegraphics[width=0.45\textwidth,clip]{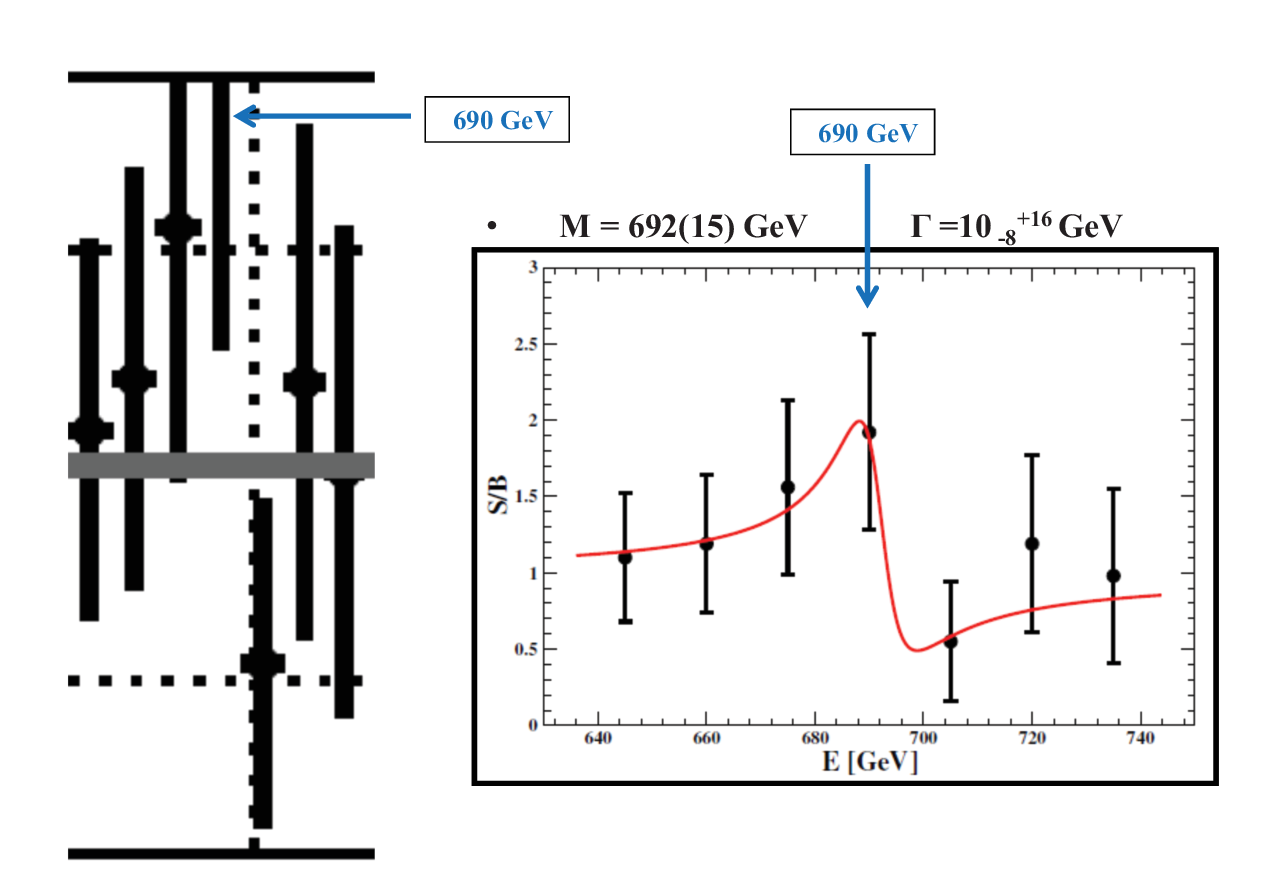}
\caption{The CMS 4-lepton data \cite{CMS_4leptons_2024} for $S/B$
at E= 640$\div$740~GeV and our fit with eq.~(\ref{sigmat}).}.
\label{CMS_7data}
\end{figure}
\begin{figure*}[ht]
\centering
\includegraphics[width=0.60\textwidth,clip]{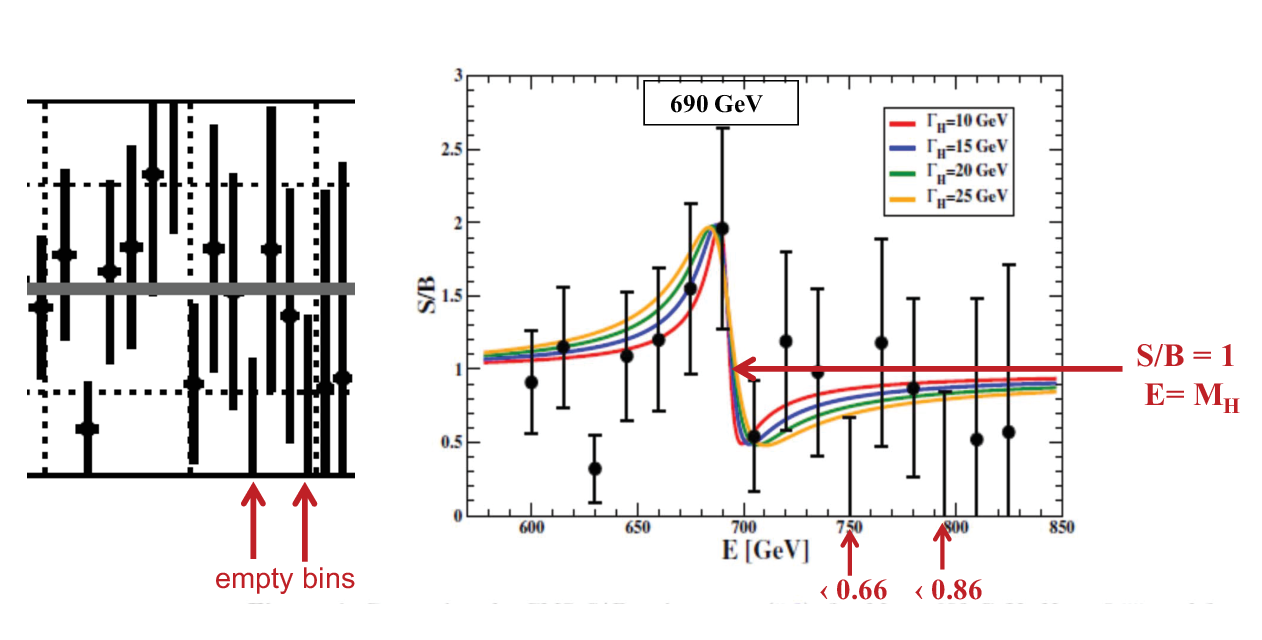}
\caption{The CMS 4-lepton data \cite{CMS_4leptons_2024} for the $S/B$
ratio from 600 to 800~GeV, as well as several curves for $M_H=692$~GeV and
four different widths. Note the two empty bins at 750 and 795~GeV, which put
the upper limits at $S/B<0.66$ and $S/B<0.86$, respectively.}.
\label{CMS_14data}
\end{figure*}
This type of excess-defect pattern, produced by the change of sign of the
interference past a BW peak, is indeed supported by the 4-lepton cross
sections of fig.~\ref{ATLAS_cross_section}, with a combined significance of
$3\sigma$; see the last column of table~\ref{leptonxsection}.
As an additional check, we have also considered the distribution of the
individual 4-lepton events. To this end, the ATLAS Collaboration has
performed a sophisticated Multivariate Analysis (MVA) \cite{atlas4lHEPData} of
the ggF-like production mode, which, depending on the degree of contamination
with the background, divides the events into four mutually exclusive
categories: MVA-high-4$\mu$, MVA-high-2e2$\mu$, MVA-high-4e, and MVA-low, with
the latter containing all three combinations. The four sets, extracted from the
corresponding HEPData file \cite{atlas4lHEPData}, are reported in
table~\ref{ATLAS-MVA}, where the original bins of 30~GeV are now grouped into
larger bins of 60~GeV, just as for the cross-section data around 700~GeV,
for which a binning of 45, 55, and 80~GeV was adopted by ATLAS. The total
number exhibits the same excess/defect sequence as for the cross-section data
in table~\ref{leptonxsection}, but the excess around 680~GeV is mainly due to
the statistically dominant MVA-low category. Nonetheless, the MVA-high events
exhibit the same substantial reduction around 740~GeV, as compared to the
expected background, thus confirming the importance of considering both
indicators (excess and defect).
The combined significance of the deviations in table~\ref{ATLAS-MVA}, at 680
and 740~GeV, is thus at the level of $4\sigma$.
Finally, since the four categories of events are mutually exclusive but were
obtained by different selection acceptances, we perform a fit to the dominant
sample of MVA-low events, which represent a homogeneous set; see
fig.~\ref{ATLAS_ggF_low}. The results of the two types of fit, viz.\ to
cross-section data and MVA-low events, are $M= 677^{+30}_{-14}$~GeV,
$\Gamma= 21^{+28}_{-16}$~GeV,
$\sigma_{\rm peak}(pp\to H \to 4l)=0.40^{+0.62}_{-0.34}$~fb, and
$M=706\,(25)$~GeV, $\Gamma=29\pm 20$~GeV,
$\sigma_{\rm peak}(pp\to H \to 4l)=0.23^{+0.28}_{-0.17}$~fb, respectively,
being both consistent with our expectations.

Further evidence of a resonance around 690~GeV can be obtained from
fig.~\ref{CMS_7data}, which shows the --- still preliminary --- CMS
4-lepton data for the $S/B$ ratio \cite{CMS_4leptons_2024} (left panel)
and our fit with eq.~(\ref{sigmat}).
To have a more complete idea of the agreement with the CMS data, we
also enlarge the energy range from 600 to 800~GeV. The data for the
$S/B$ ratio are then presented in fig.~\ref{CMS_14data}, together with
various curves for the same mass $M_H= 692$~GeV and different
widths.  Note the two empty bins at 750 and 795 GeV, with their error bars
representing the CMS estimates for the upper limits that one could expect with
more statistics, namely $S/B<0.66$ and $S/B<0.86$, respectively. Together with
the very low point at 705~GeV, and in view of the large error bars of the
remaining points, this means that, in the region above 690~GeV, values with
$S/B$ considerably smaller than unity have a large probability content. 

In fact, if we compute the ratio between the total number of
observed 4-lepton events, in the range 705$\div 825$~GeV, and the total
number of expected background events, we find an average
$\langle S/B\rangle=0.68\,(17)$.
This is in excellent agreement with the analogous determination
$\langle S/B\rangle=0.69\,(18)$ obtained from the 14 events observed by ATLAS,
for $710\div830$~GeV, when compared to the expected 20.3 events (see
table~\ref{ATLAS-MVA}). The theoretical curves, especially those of green and
yellow colour for widths $20\div25$~GeV, can thus provide a clue with their
prediction of a slow increase in $S/B$ from about 0.5 at 700~GeV up to about
0.8 at 800~GeV.
\scriptsize
\begin{table}[!b]
\caption{The 95$\%$ upper limits $\sigma^{\rm obs}$ to the cross section
$\sigma(pp\to X \to h h)$ observed by ATLAS \cite{ATLAS_BBGG_paper}, from
the analysis of the $b\bar b + \gamma\gamma$ final state. The events are
grouped according to their invariant mass $m(\gamma\gamma jj)=E$ around each
nominal energy $E \equiv M_X$. Error bars in the observed entries take only
into account the $\sqrt{N}$ statistical uncertainty of the final
$b\bar b + \gamma\gamma$ events. In the third column, we report the
corresponding upper limits on the expected background with
$\pm 1\sigma$ and $\pm 2\sigma$ theoretical uncertainties (see the HEPData
file of \cite{ATLAS_BBGG_paper}.)}
\begin{center}
\begin{tabular}{ccc}
\hline\hline
$M_X$ (GeV) & $\sigma^{\rm obs}$ [fb] & $\sigma^{\rm expected}$ [fb] \\ \hline
600 & 73.6\,(14.3) & $81.1^{ {+43.3}^{+119.0} }_{ {-22.7}_{-37.6} }$ \\ \hline
650 & 149.3\,(20.3)& $84.4^{ {+44.4}^{+120.1} }_{ {-23.6}_{-39.1} }$ \\ \hline
700 & 49.4\,(12.0) & $76.5^{ {+40.0}^{+109.6} }_{ {-21.4}_{-35.4} }$ \\ \hline
750 & 44.5\,(12.0) & $71.7^{ {+37.6}^{+103.3} }_{ {-20.0}_{-33.2} }$ \\ \hline
800 & 71.0\,(14.0) &  $65.8^{ {+35.1}^{+96.5} }_{ {-18.4}_{-30.5} }$ \\ \hline
\end{tabular}
\end{center}
\label{sigmas}
\end{table}
\normalsize
\begin{figure*}[ht]
\centering
\includegraphics[width=1.0\textwidth,clip]{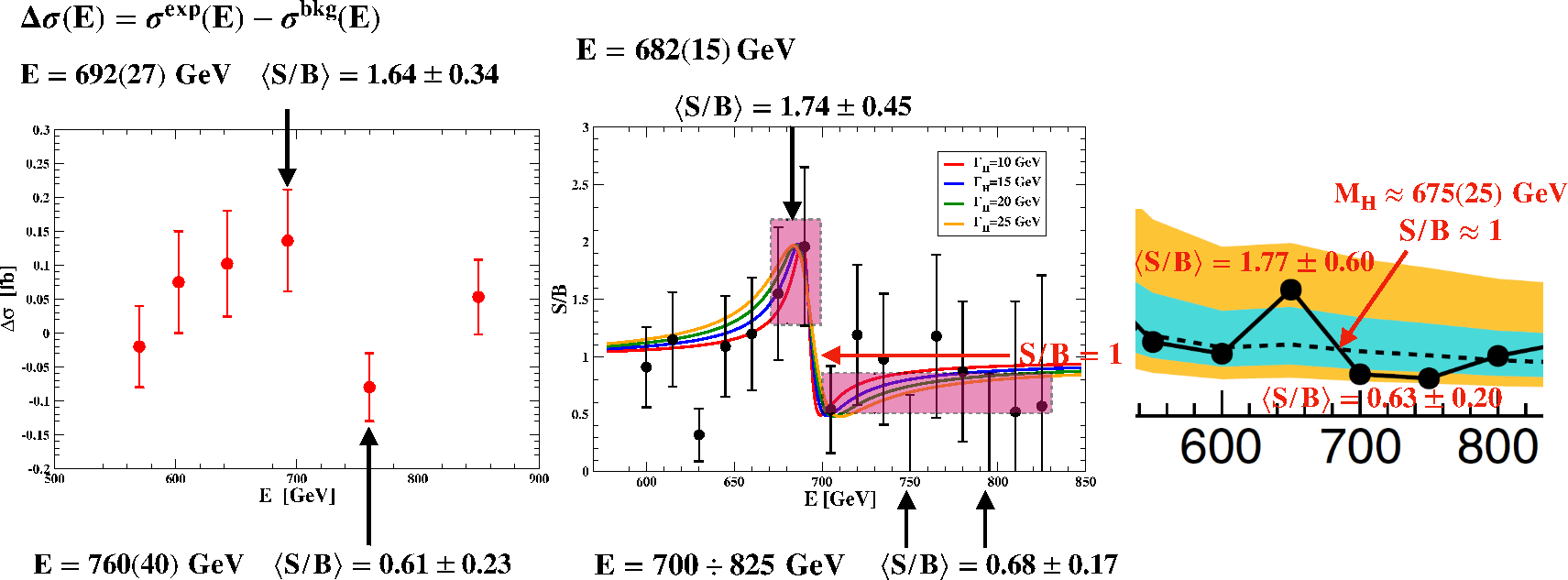}
\caption{The ATLAS 4-lepton data of fig.~\ref{ATLAS_cross_section},
together with the CMS data of fig.~\ref{CMS_14data} and the analogous ATLAS
data \cite{ATLAS_BBGG_paper} for $\sigma(pp\to X\to hh)$ as extracted from
the $b\bar b + \gamma\gamma$ final state.}.
\label{3-data}
\end{figure*}

Let us finally consider the $H \to hh$ decay, where, for
$B(H\to hh)=0.05\,(1)$, we would expect
$\sigma_{\rm peak} (pp\to H\to hh)=50\,(10)$~fb. This prediction can be
compared with the ATLAS search \cite{ATLAS_BBGG_paper} for a new resonance
$X$ through the analysis of the process
$pp\to X \to hh \to b\bar b + \gamma\gamma$. The results around 700~GeV are 95$\%$ upper bounds 
for the cross section $\sigma(pp\to X \to hh)$ and are
reported in table~\ref{sigmas}. These cross sections are obtained
from the distribution of events with invariant mass $m(\gamma\gamma jj)=E$
around each nominal energy $E \equiv M_X$. The analysis is performed in the
narrow-width approximation (10 MeV) and does not account for any
interference with the background. 

From table~\ref{sigmas} one gets the impression of a modest $+1.2\sigma$
excess at 650~GeV, followed by two slight $-1.3\sigma$ defects. 
However, this is not the right perspective. The point is that, in our mass
region and to a very good approximation, the theoretical uncertainties simply
shift the central values up and down. Therefore, if we compute the difference
between the observed values at 650 and 700~GeV, which is $99.9\,(23.6)$~fb,
the result is much larger than the corresponding differences between the
central values $(84.4 -76.5)$~fb $=7.9$~fb or along the $+1\sigma$ and
$+2\sigma$ contours, i.e., $(128.8-116.5)$~fb $=12.3$~fb and
$(204.5-186.1)$~fb $=18.4$~fb, respectively. Understanding this sizeable
discrepancy thus requires an asymmetric effect that increases the number of
events with $m(\gamma\gamma jj)$ around 650~GeV and, at the same time, lowers
the corresponding number with $m(\gamma\gamma jj)$ around 700~GeV, like with a
resonance between 650 and 700~GeV. Even though the observed entries express
upper bounds, so that, strictly speaking, they cannot be fitted with
eq.~(\ref{sigmat}), to roughly estimate $M_H$ and $\Gamma_H$ we can nonetheless try to
understand the orders of magnitude. In this case, for $M_H\sim675$~GeV,
$\Gamma_H \sim 25$~GeV, $\sigma_{\rm peak}\sim  50$~fb, and an average
background cross section $\sigma_B\sim 80$~fb, we find
$\sigma_T(650)\sim140$~fb and $\sigma_T(700)\sim40$~fb, which numbers are
close to the observed limits.

As a final evidence, we put together three sets of data in
fig.~\ref{3-data}. Namely, the ATLAS 4-lepton cross sections of
fig.~\ref{ATLAS_cross_section}, the CMS $S/B$ ratios of fig.~\ref{CMS_14data},
and the ATLAS limits \cite{ATLAS_BBGG_paper} for 
$\sigma(pp\to X\to hh)$ from the $b\bar b + \gamma\gamma$ final state. It is
evident that the three sets indicate the same type of excess/defect pattern,
with a combined statistical significance that is far from negligible.

\section{Conclusions}
With the phenomenology expected for the second resonance, the dominant decay
process $H \to t \bar t$ has a background that is about 100 times larger than
the expected signal. In this situation, the slight $1\%$ excess seen by ATLAS,
in our mass region around 675~GeV, is already a very interesting indication,
even though we are only speaking of a $1\sigma$ effect. Concerning the other
two main channels $H\to ZZ$ and $H\to hh$, the BW peak is comparable to the
background. We have thus exploited the expected signature with the simple
simulation reported in table~\ref{simulation}. This indicates the importance
of sequences with excess/defect of events. This type
of pattern is indeed observed in the ATLAS and CMS 4-lepton data
(see figs.~\ref{ATLAS_cross_section}--\ref{CMS_14data}) and in the ATLAS
$(b\bar b + \gamma\gamma)$ final state (see the third panel of
fig.~\ref{3-data}). Of course, this is just a data sample, as compared to the
large number of LHC measurements. However, the final states we have considered
show a good resolution in invariant mass, so that our indications should be
taken seriously. Likewise, we can also consider other data, for instance the
CMS search for heavy resonances $X$ through the chain $X \to hh\to b\bar b WW$.
From fig.~19 (upper panel) of \cite{CMS_X_bbWW}, the $S/B$ ratio is seen to
decrease from about 1.5 at 600~GeV down to about 0.5 at 750~GeV. This sizeable
$2\sigma$ defect would be consistent with the trend observed, by both ATLAS
and CMS in the 4-lepton channel, and by ATLAS in the
$b \bar b +\gamma\gamma$ final state, in the same region of invariant mass.  

Concerning the other important decay channel into a pair of charged $W$s,
we have considered the $H\to WW \to 2l 2\nu$ process. As explained, the second
resonance is essentially produced via the ggF mechanism. Therefore, when
comparing with the existing CMS measurements \cite{CMS_X_WW_PAS}, the second
resonance is in the class of models where the VBF production mode is
irrelevant. This is the case $f_{\rm VBF}= 0$ in fig.~4 (top left) of
\cite{CMS_X_WW_PAS}. From the numbers reported before, i.e., a partial width
$\Gamma(H\to WW)\sim 3.3$~GeV and a total width
$\Gamma(H\to {\rm all})\sim 25\div 35$ GeV, we find a branching ratio
$B(H\to WW)\sim  0.11\,(2)$. Therefore, for a ggF production cross section of
about 1~pb, we expect a resonant contribution
$\sigma(pp\to H\to WW\to 2 l 2\nu)\sim 5\,(1) \times 10^{-3}$~pb, well
consistent with the CMS 95$\%$ upper limit of $0.02\div 0.03$~pb around
700~GeV. Analogous considerations could be applied to other samples of data
in which the weakness of the expected signal and/or the low statistics do
not allow for stringent tests.

But also other indications were previously collected in \cite{universe, LHEP, EPJC}. For
instance, the ATLAS inclusive $\gamma\gamma$ events \cite{atlas2gammaplb}
indicate a local $3\sigma$ excess at $684$~GeV, and a fit to these data with
eq.~(\ref{sigmat}) yields a resonance mass $M_H=696\,(12)$~GeV. Analogously,
the CMS-TOTEM $\gamma\gamma$ events produced in $pp$ diffractive scattering
\cite{PRD_CMS_TOTEM} indicate another local excess of $3\sigma$ in the region
of invariant mass $M_H=650\,(40)$~GeV. Since these deviations from the
background lie in our theoretical mass region, it appears unjustified to
invoke the so called ``look-elsewhere effect'' so as to downgrade their
statistical significance. Therefore, if we count all local deviations and
start to sum up the squares of the various sigmas, we could easily reach
the traditional $5\sigma$ level. However, we believe that this may be
premature. The main message of our present work is that the proposed new
resonance requires a dedicated search in which both excesses and defects of
events should be considered in order to determine the overall statistical
significance. The whole idea of just looking for ``bumps'' is totally
inadequate here and a p-value based on this naive criterion may be
dramatically wrong.

\end{document}